\documentclass[useAMS,usenatbib]{mn2e}
\usepackage{latexsym,graphicx,natbib}

\def\simless{{\th \rlap{\raise 0.5ex\hbox{$\scriptstyle  {<}$}}
    {\lower 0.3ex\hbox{$\scriptstyle  {\sim}$}} \th }}  
\def\simgreat{{\th \rlap{\raise 0.5ex\hbox{$\scriptstyle  {>}$}}
    {\lower 0.3ex\hbox{$\scriptstyle  {\sim}$}} \th }}  
\def\greateq{{\th \rlap{\raise 0.5ex\hbox{$\scriptstyle  {>}$}}
    {\lower 0.3ex\hbox{$\scriptstyle  {-}$}} \th }}  
\def\lesseq{{\th \rlap{\raise 0.5ex\hbox{$\scriptstyle  {<}$}}
    {\lower 0.3ex\hbox{$\scriptstyle  {-}$}} \th }}  

\def\th{\thinspace}
\def\ts{{\raise 0.3ex\hbox{$\scriptstyle {\th \sim \th }$}}}
\def\be{\begin{equation}}
\def\ee{\end{equation}}
\newcommand{\ltaraw}{$\; \buildrel < \over \sim \;$}
\newcommand{\lta}{\lower.5ex\hbox{\ltaraw}}
\newcommand{\gtaraw}{$\; \buildrel > \over \sim \;$}
\newcommand{\gta}{\lower.5ex\hbox{\gtaraw}}
\newcommand{\mdot}{M_{\odot}}

\newcommand{\msun}{{\rm\,M_\odot}}

\newcommand{\apj}{ApJ}
\loadboldmathitalic 
\title[Optical emission in ULX binary systems]
{Optical emission from massive donors in ULX binary systems}
\author[Patruno \& Zampieri]
{
Alessandro Patruno$^1$ \& Luca Zampieri$^2$ \\
$^1$Astronomical Institute `A. Pannekoek', Univeristy of Amsterdam,
Kruislaan 403, Amsterdam, The Netherlands: \tt{apatruno@science.uva.nl}\\
$^2$INAF - Osservatorio Astronomico di Padova, Vicolo
dell'osservatorio 5, 35122 Padova, Italy :{ \tt luca.zampieri@oapd.inaf.it}} 
\date{\today}

\def\simless{\mathbin{\lower 3pt\hbox
   {$\rlap{\raise 5pt\hbox{$\char'074$}}\mathchar"7218$}}} 
\def\simgreat{\mathbin{\lower 3pt\hbox
   {$\rlap{\raise 5pt\hbox{$\char'076$}}\mathchar"7218$}}} 

\def\ergs{{\rm\,erg\,s^{-1}}}

\def\msun{M_\odot}

\def\be{\begin{equation}}
\def\ee{\end{equation}}
\def\baray{\begin{eqnarray}}
\def\earay{\end{eqnarray}}

\def\gcm3{{\rm g\,\, cm^{-3}}}

\begin{document} 
\maketitle 
\begin{abstract}

We present evolutionary tracks of binary systems with high mass
companion stars and stellar-through-intermediate mass BHs. Using
Eggleton's stellar evolution code, we compute the luminosity produced
by accretion from the donor during its entire evolution. We compute
also the evolution of the optical spectrum of the binary system taking
the disc contribution and irradiation effects into account. The
calculations presented here can be used to constrain the properties of
the donor stars in Ultraluminous X-ray Sources by comparing their
position on the HR or color-magnitude diagrams with the evolutionary
tracks of massive BH binaries. This approach may actually provide
interesting clues also on the properties of the binary system itself,
including the BH mass.  We found that, on the basis of their position
on the color-magnitude diagram, some of the candidate counterparts
considered can be ruled out and more stringent constraints can be
applied to the donor masses.

\end{abstract}
\begin{keywords} 
galaxies: M81, NGC1313, NGC4559, Holmberg II --- X-rays: binaries --- X-rays: galaxies
\end{keywords} 

\section{Introduction}\label{introduction}

Point like off nuclear X-ray sources with luminosities well in excess
of the Eddington limit for a stellar mass black hole, have been
discovered in a large number of nearby galaxies (e.g.~\citealt{2002ApJS..143...25C}; \citealt{sgtw04}; \citealt{2005ApJS..157...59L}). These Ultraluminous
X-ray Sources (ULXs) are too dim to be low luminosity AGNs and too
bright to be normal X-ray binaries (XRBs) emitting below the Eddington
limit. In this paper we define a ULX as a source with bolometric
luminosity in excess of the Eddington limit for a stellar mass black
hole of $20\msun$ and less luminous than a few $10^{41}\rm erg/s$.
This definition implies that a Galactic XRB radiating isotropically at
or below the Eddington limit can not be a ULX.

In spiral or starburst galaxies ULXs turn out to be associated with
star forming regions, emission nebulae and stellar clusters (\citealt{zfrm02},
\citealt{pm02}). These facts along with, in some
case, the detection of stellar optical counterparts
 (\citealt{2001MNRAS.325L...7R}; \citealt{2002MNRAS.335L..67G}; \citealt{2002ApJ...580L..31L}; \citealt{2004ApJ...602..249L};\citealt{kwz04};\citealt{zamp1}; 
\citealt{2005ApJ...629..233K}; \citealt{mucetal05,
mucetal07}; \citealt{scpmw04}) strongly indicate an association
between ULXs and young massive stars, although the nature of the
accreting compact object remains unclear.

A certain number of ULXs, however, appear as isolated X-ray sources
with no obvious counterpart at any wavelength (\citealt{sm04}) and
without a clear association with star forming regions or emission
nebulae.
Some low-luminosity ULXs may possibly be present also in elliptical
galaxies (\citealt{jcbg03}), but their nature and the actual
evidence for their existence is not well established (see e.g. \citealt{iba04},
\citealt{aglc04}, \citealt{2005ApJ...622L..89G}).

The nature of the ULXs in spiral galaxies is less controversial.
Different models have been proposed to explain the large luminosities
reached by these sources. One of the favored models consists of an
intermediate mass black hole (IMBH) with a mass in the range
$10^{2}-10^{3}M_{\rm \odot}$, accreting from an high mass donor
star. The presence of an IMBH can account for most of the
observational properties of ULXs in a rather straightforward way. For
instance, the observed cool disc spectra of some ULX can be explained
with the fact that the innermost stable circular orbit of an IMBH is
larger than that of a stellar mass black hole (e.g. \citealt{mfm04}). 
The detection of a $\sim$50-160 mHz quasi periodic
oscillations in the power density spectrum of M82 X-1 and NGC5408 X-1
(\citealt{2003ApJ...586L..61S}, \citealt{2004ApJ...614L.113F}, 
\citealt{2006MNRAS.365.1123M}; \citealt{stroh07}), 
the very high luminosity of some ULXs
($\sim 10^{41} \ergs$) along with their cool discs, and the energy
content and morphology of the nebulae around some of them (\citealt{pm02}) all suggest an IMBH interpretation. The main problem
with this interpretation resides in the formation mechanism of such an
extreme object. In fact, if IMBHs with masses in excess of $\sim 100
M_\odot$ exist, they will require a new formation root with respect to
the stellar black holes in our Galaxy and to the supermassive black
holes in Active Galactic Nuclei. Until now two scenarios have been
proposed to form a black hole in the intermediate mass range: the
runaway collision of massive stars in dense open clusters (\citealt{2004Natur.428..724P}, \citealt{2004ApJ...604..632G}) and the primordial
collapse of a very high mass star with zero metallicity (\citealt{2000ApJ...540...39A},
 \citealt{2001ApJ...551L..27M}). Both the mechanisms however suffer of a
certain degree of uncertainty related to the incomplete knowledge of
the behavior of very massive stars. Therefore we have no final
evidence that an IMBH can really form. Furthermore, the interpretation
of the soft components observed in some ULXs in terms of cool
accretion discs is not univocal (e.g. \citealt{2005ApJ...631L.109C}, 
\citealt{2005ApJ...635..198D}, \citealt{2005MNRAS.357.1363R}, \citealt{fk06}, 
\citealt{gonc06}, \citealt{2006MNRAS.368..397S}).

Other interpretations in terms of stellar (or quasi-stellar) mass
black holes have been proposed. A mechanical (\citealt{kdwfe01}) or a
relativistic beaming (\citealt{kfm02}) can reproduce
the observed luminosities of ULXs up to a few $10^{40}\ergs$ with a
beaming factor around $\sim 10.$ At most, as suggested by 
\citealt{2005MNRAS.357..275K}, accretion from helium rich matter from a geometrically
thick disc can generate luminosities up to $\sim 5\times10^{40}\ergs$.
However, luminosities in excess of $5\times 10^{40}\ergs$ ($\sim 5$\%
of the ULX population) and the isotropy of the ionized nebulae around
some ULXs can not be easily explained in terms of beaming models. On
the other hand, photon bubbles disc instabilities (\citealt{b02}, \citealt{b06}) and
emission from a slim disc (\citealt{wmm01}, \citealt{ezkmw03})
can produce genuine isotropic super-Eddington luminosities around 10
times the Eddington limit.

As shown by \cite{2005MNRAS.356..401R}, a normal binary
with a stellar mass black hole and a donor star with initial mass
$\simgreat 10\mdot$, can in principle explain a large sample of ULXs
if a super Eddington luminosity around $10$ is allowed (see also
\citealt{2003MNRAS.341..385P}). However, the typical
temperature of a slim disc in this regime ($\sim$ 1--2 keV) is
inconsistent with the observation of some cool disc sources which are,
at the same time, the most luminous ULXs. Furthermore, luminosities in
excess of a few $10^{40}\ergs$ are not attainable with slim discs
around stellar mass black holes. On the other hand, the photon bubble
instability model makes no predictions on the observable accretion
disc temperature and therefore cannot be compared directly with
observations.

Thanks to the high precision astrometry of the {\it Chandra} X-ray
observatory, we know that some ULXs have optical counterparts which
match with the X-ray source position (\citealt{2002ApJ...580L..31L}, 
\citealt{2004ApJ...602..249L}, \citealt{zamp1}, \citealt{kwz04}, 
\citealt{sm04}, \citealt{2005ApJ...629..233K},
\citealt{mucetal05, mucetal07}, \citealt{Liu07}). Most
of them are suspected to be main sequence (MS) high mass stars or
supergiant stars of uncertain spectral type. The ambiguity arises
because, until now, optical spectra of these stars either are not
available, or they are too noisy and with peculiar spectral features
(\citealt{2004ApJ...602..249L}), or there is more than an optical counterpart in the
X-ray error box (\citealt{sm04}, \citealt{mucetal05}).

In this paper, we present the evolutionary tracks of binary systems
with high mass companion stars and stellar-through-intermediate mass
BHs and compare them with the properties of the donor stars in ULX
binary systems. In \S 2 we present the model adopted to evolve the
binary system. In \S 3, we summarize the properties of the four ULXs
considered and their optical counterparts. Our results are
presented in \S 4 and compared with the properties of ULX counterparts
in \S 5. Conclusions follow in \S 6.

\section{Model}

We consider a binary system with a stellar mass BH (10 $M_\odot$) or
an IMBH ($\ga 100 M_\odot$) and a companion star that is orbiting
around it and eventually transferring matter onto it. The evolution of
the binary is computed using an updated version of the Eggleton code
(\citealt{1971MNRAS.151..351E}; \citealt{1995MNRAS.274..964P}). 
The calculation follows that
presented in \cite{patruno}. We assume a Population I chemical
composition (Y=0.28, Z=0.02) and allow for non-conservative stellar
evolution, taking into account wind loss from luminous stars (\citealt{1988A&AS...72..259D}). 
The adopted mixing length parameter and overshooting
constant are $\alpha=2.0$ and $\delta_{ov}=1.2$, respectively (\citealt{1998MNRAS.298..525P}). 
Loss of angular momentum is also accounted for through
emission of gravitational waves (see e.g. \citealt{ll75}) and
particles in winds (\citealt{spvdh97}).

If accretion occurs via Roche-lobe overflow (RLOF), we assume that an
accretion disc will form. In this work we adopt an efficiency
$\eta=0.1$ for the conversion of gravitational potential energy into
radiation in a disc. The bolometric luminosity of the X-ray source is
$L=\eta\dot{M}c^{2}$, where $\dot{M}$ denotes the average mass
transfer rate from the donor as computed from the numerical code.

In case accretion occurs via mass loss from a stellar wind (wind-fed
accretion; WFA), the situation is somewhat different if one considers
stellar mass BHs or IMBHs, as the latter are bigger and have a
stronger gravitational potential well. Hence, owing to the larger
gravitational capture radius, a comparatively larger fraction of the
wind emitted by the donor star can be captured at corresponding
orbital separations.  Furthermore, a large fraction of the accreting
particles have a significant orbital angular momentum. This might be
sufficient to form an accretion disc even in a wind-fed system with a
IMBH, although further investigations are required to completely
understand the geometry of these discs. In the following we will limit
our analysis only to WFA systems with IMBHs. For the WFA model we
adopt a modified version of the RLOF model, adding an equation to
compute the WFA accretion rate (as described in \citealt{patruno}). An accretion disc is assumed to form also in this case but,
unless the donor is very massive ($\simgreat 60 \msun$, not considered
in this work), the accretion rate is essentially negligible.

\subsection{Initial orbital separation}

The evolution of a binary system containing a BH/IMBH depends on the
initial orbital separation. The companion star can fall in three
different zones, defined in terms of the zero-age/terminal-age main
sequence (ZAMS/TAMS) radius of the donor $R_{\rm ZAMS}$/$R_{\rm TAMS}$, the tidal
radius of the system $R_{\rm T}$ and the Roche lobe radius of the
companion $R_{\rm L}$. The expressions of these radii are (\citealt{1983ApJ...268..368E}, \citealt{dk91}):
\begin{eqnarray}
&& R_{\rm ZAMS}=\left( \frac{M}{M_{\odot}} \right)^{0.57}R_{\rm \odot}\\
&& R_{\rm TAMS}=1.6\left(\frac{M}{M_{\rm \odot}} \right)^{0.83}R_{\rm \odot}\\
&& R_{\rm T}=\left(\frac{M_{\rm BH}}{M}\right)^{1/3}R\\
&& R_{\rm L}=\frac{0.49q^{2/3}a}{0.6q^{2/3}+\ln(1+q^{1/3})} \, ,
\end{eqnarray}
where $M$ and $R$ refer to the donor star, $M_{\rm BH}$ is the black
hole mass, $q=M/M_{BH}$ is the mass ratio, and $a$ is the orbital
separation.

In a reference frame centered on the BH/IMBH, the \textbf{first zone}
lies between $a=0$ and $a\sim R_{\rm T}$. If the companion star enters
in the first zone, it is disrupted by the strong tidal forces of the
BH, likely resulting in an outburst of short duration (\citealt{1988Natur.333..523R},
\citealt{u99}, \citealt{alp00}). This possibility is not
considered in the present work.

The \textbf{second zone} is defined as the interval of orbital
separations where the stellar surface can reach contact with the Roche
lobe radius $R_{\rm L}$ sometime during MS. 
In the following we will refer to the binaries whose donor falls in
the second zone as ``case A'' systems. In order for the initial
orbital separation to be in this range, either the companion star is
tidally captured by the BH/IMBH (\citealt{hpza04}, \citealt{2006MNRAS.372..467B}) or the binary system undergoes an
exchange interaction with another star (\citealt{2004ApJ...613.1143B},
\citealt{al04}, \citealt{2006ApJ...642..427B}).

If the initial orbital separation $a$ is such that $R_{\rm L}>R_{\rm TAMS}$, the companion is in the
\textbf{third zone}. In this zone, mass transfer occurs through RLOF 
only when the donor is in the H-shell phase (case B) or during the 
He-shell burning (case C). If the star falls in
this zone, during the main sequence the only possible accretion mechanism
is the gravitational capture of wind particles. 
WFA IMBHs systems can reach ULX luminosities even if the
orbital separation is a few astronomical units, thanks to the enhanced
gravitational focusing effect of the IMBH. However, for larger
separations, the source is fainter and has a mean luminosity typical
of bright Galactic XRBs (see \citealt{patruno} for an extended
discussion). If the initial separation is very large (usually several
tens of astronomical units), neither RLOF nor WFA are possible, and
the system evolves essentially as a detached, non-interacting binary.

\subsection{Stability of the accretion disc around an IMBH}\label{inst}

The accretion disc around a BH is steadily fueled if the mass transfer
rate from the donor exceeds the critical mass transfer rate (\citealt{dlhc99}):
\begin{eqnarray}\label{crit}
\dot{M}_{cr}=2.4 \cdot 10^{-6}\Bigg(\frac{M_{BH}}{10^2 \, M_{\odot}}\Bigg)^{1/2}
\Bigg(\frac{M}{15M_{\odot}}\Bigg)^{-0.2}\nonumber\\
\times\Bigg(\frac{M_{\rm BH}+M}{10^2 \, \msun}\Bigg)^{-0.7}
\Bigg(\frac{a}{{\rm 1 \, AU}}\Bigg)^{2.1} M_{\odot} \, {\rm yr}^{-1}
\end{eqnarray}
The critical mass accretion rate given above takes already into
account the stabilizing effect of the increase in temperature due to
the disc self-irradiation. In our model, all the IMBH systems
accreting through RLOF turn out to be persistent if the donor mass is
higher than $\sim 10M_{\odot}$, in agreement with the findings of
\cite{pzdm} and \cite{patruno}. For a stellar mass BH the stability of the disc is
guaranteed down to donor masses $\sim 5\msun$.

Assuming that equation~[\ref{crit}] holds also for WFA discs around
IMBHs, thermal instabilities are likely to arise. In fact, the mass
transfer rate in WFA systems is smaller than that in RLOF systems. All
the IMBH WFA systems with a donor star lighter than 30$M_{\odot}$
might experience transient accretion (\citealt{patruno}). 
Stars heavier than 30$\msun$
likely produce a transient phase during the early MS (when the mass
transfer rate is small) and a persistent phase during the late MS
(when the mass transfer is higher than the critical mass transfer
rate). So we expect rapid outbursts and quiescent states (similar to
those observed in Galactic transient XRBs) in low mass companion WFA
systems and, in general, during the early evolution of WFA
systems. The recurrence time and duty cycle are difficult to
estimate. As in a dynamical exchange interaction it is highly probable
that the companion star falls in the \textbf{third zone}, WFA systems
might be rather common.  Thus, a significant population of IMBH
binaries that appear as transient ULXs or even transient XRBs might
exist in external galaxies or even in our own galaxy. Some sources of
this transient ULX population start to be identified
(e.g. \citealt{fk07}).

\subsection{X-ray reprocessing in ULXs}

An important effect to consider when assessing the properties of ULX
counterparts is the possible strong optical-UV contamination caused by
the emission of the accretion disc itself and, in case of isotropic
emission, also by the reprocessed X-ray radiation. The main
reprocessing sites are the outer part of the accretion disc and the
donor surface. We follow the evolution of the optical luminosity and
colors of the binary system taking irradiation effects into account.

Our calculation relies on the same assumptions discussed in
\cite{2005MNRAS.362...79C} and \cite{mucetal07}. More specifically, a
standard Shakura-Sunyaev disc (e.g. \citealt{frakira})
is assumed and both the X-ray irradiation of the companion and the
self-irradiation of the disc are accounted for. However, in order to
keep our treatment simple, the companion star is taken to be spherical
and at uniform temperature, neglecting the effects produced by the
Roche lobe geometry and also those related to the (possible)
deformation induced by radiation pressure.
We adopt a simplified description of radiative transfer for the
interaction of the X-rays with the disc and donor surfaces (a X-ray
illuminated plane-parallel atmosphere in radiative equilibrium) and do
not include limb and gravity darkening. The shadowing of the disc on
the star is continuously monitored during the whole evolution and
never overcomes $\sim$ 15\%.

The computed luminosity and colors depend on the masses of the donor
and BH, the binary period (or orbital separation), the accretion rate
and the (unirradiated) temperature of the donor, in addition to the
inclination angle $i$ and the orbital phase $\phi$ that are kept fixed
at $\cos i=1$ (disc face-on) and $\phi=0$ (superior conjunction). The
accretion efficiency and the albedo of the donor surface layers were
chosen to be 0.17 and 0.9 respectively. Following
\cite{2005MNRAS.362...79C}, we took the hardness ratio $\xi=F_X(<1.5\,
{\rm keV})/F_X(>1.5\, {\rm keV})=0.1$. The absorption parameters in
the same two spectral bands were selected as $k_s=2.5$ and $k_h=0.01$.
The R, V and B magnitudes of the (irradiated) disc plus donor have
been computed using the output values of the parameters provided by
the binary evolution code. In particular, the accretion rate is
instantaneously taken to be equal to the mass transfer rate from the
companion. When the accretion rate overcomes $\dot{M}_{Edd}$, we
impose $\dot{M}=\dot{M}_{Edd}$ and assume that the excess mass is
expelled from the system.

\section{ULXs with optical counterparts}

The purpose of our work is to provide a tool for comparing the
observational properties of ULXs with identified optical companions
with those predicted for the donors of our BH/IMBH binary model.  As
reference cases, we consider four ULXs for which sufficient
information on the optical counterparts are available to allow a
meaningful comparison.
They are: NGC1313 X-2 (\citealt{mfmf03}, \citealt{zamp1},
\citealt{mucetal05, mucetal07}, \citealt{Liu07}), Holmberg II X-1
(\citealt{dmgl04}, \citealt{kwz04}), NGC4559 X-7 (\citealt{csmwmp04},
 \citealt{scpmw04}) and M81 X-9 (\citealt{mfm04}). 
None of the ULXs considered have a unique optical counterpart.
In the case of NGC1313 X-2, there are two optical counterparts in the {\it Chandra} error
box, identified on {\it VLT} and {\it HST} images.
Photometry and modeling of the donor emission are consistent with the
following interpretation: an early B MS star of $10-20\msun$ and a
red/yellow supergiant of $10\msun$ (\citealt{mucetal05},
\citealt{mucetal07}). For the three other ULXs, only optical
photometry is available, and the nature of the donor star is still
uncertain.

Fits of the X-ray spectra of these ULXs have been carried out by
several authors. Parameters of the best fit obtained with different
models are listed in Table~\ref{tab1} and are used to obtain an
estimate of the total X-ray luminosity of the system, extrapolating
the flux in the energy range $0.05-20\,\rm keV$ using the web
interface to PIMMS (WebPIMMS v. 3.9c). 
For M81 X-9 we adopt the bolometric luminosity reported by \cite{mfm04}.

\begin{table*}
\begin{center}
\caption{Spectral parameters and luminosity of the ULXs considered in this work.}
\label{tab1}
\begin{tabular}{llllllll}
\hline\hline
ULX  & $N_{H}$ & $kT$ & $\Gamma$ &  $L_{X}$ & $L_{bol}^a$  & Model$^b$ & source \\ \hline
 & $(\rm 10^{21} cm^{-2})$ & $(eV)$ & & $(10^{39} \rm erg\,s^{-1})$ & $(10^{39} \rm erg\,s^{-1})$ &  & \\ \hline
NGC1313 X-2 & $3.13^{+0.92}_{-0.37}$ & $200^{+40}_{-50}$ & $2.23^{+0.15}_{-0.09}
$ & $3.4$ & $7.0$ & PL+MCD & \cite{zamp1} \\
Holmberg II X-1  & $1.4^{+0.3}_{-0.03}$ & $128^{+22}_{-13}$ & $2.40^{+0.07}_{-0.08}$ & $17$ & $37.7$ & PL+BB & \cite{dmgl04}\\
NGC4559 X-7 & $4.3^{+0.9}_{-1.1}$ & $120^{+10}_{-10}$ & $2.23^{+0.06}_{-0.05}$ & $19$ & $36.9$ & PL+BB & \cite{csmwmp04} \\
M81 X-9 & $2.3$ & $260^{+20}_{-50}$ & $1.73$ & $11$ & $27$ & PL+MCD & \cite{mfm04} \\
\hline 
\end{tabular}  
\end{center}
$^a$Total X-ray (0.05--20 keV) luminosity computed using the web interface to PIMMS (ver. 3.9c).\par
$^b$PL=power-law; MCD=multicolor disc blackbody; BB=blackbody
\end{table*}

Finally, we briefly summarize the main properties of the optical
counterparts of the four ULXs considered in this work. All the
spectral classifications and mass estimates reported below are rather
uncertain and assume that the donor can be treated as if it were a
single star. In addition, except for NGC1313 X-2, contamination of the
optical emission from the accretion disc and X-ray irradiation at the
donor surface are not accounted for.

\begin{itemize}

\item \textbf{NGC4559 X-7} (\citealt{csmwmp04}, \citealt{scpmw04}):
six stars fall inside and two more counterparts are slightly outside
the Chandra error box of this ULX. The most luminous is a blue object
with properties consistent with a main sequence star of $\sim 20\msun$
(\citealt{2005MNRAS.362...79C}). Using color-magnitude diagrams based on the
Geneva tracks, \cite{scpmw04} suggest that the colors of the six
counterparts inside the error box are consistent with donors with mass
between 9 and 25$\msun$.

\item \textbf{NGC1313 X-2} (\citealt{zamp1}, 
\citealt{mucetal05, mucetal07}; \citealt{Liu07}):
two objects, C1 and C2, are present inside the Chandra error box of
this source having spectral type consistent with an early B MS star of
$\sim 10$--$18\msun$ or a G supergiant of $\sim 10\msun$, respectively
(\citealt{mucetal05,mucetal07}).
\cite{Liu07} find that the spectral energy distribution
of object C1 is consistent with either a $\sim 8\msun$ star of very
low metallicity or an O spectral type, solar metellicity star of $\sim
30\msun$.

\item \textbf{Holmberg II X-1} (\citealt{dmgl04}, \citealt{kwz04}): 
as other ULXs, this source is embedded in a ionized nebula
(\citealt{pm02}). A star with color $B-V$ and optical
magnitude consistent with a $O4-O5$V or $B3$Ib spectral type falls
inside the X-ray error box. The corresponding mass is, respectively,
$\simgreat 60\msun$ for a MS donor and $\sim 20-25\msun$ for a
supergiant.

\item \textbf{M81 X-9} (\citealt{mfm04}): this source is fully
embedded in a ionized nebula inside the dwarf companion galaxy of M81,
Holmberg IX (\citealt{pm02}). Inside the X-ray error box there
is a blue luminous star with magnitude $B=22.1$ that, at a distance of
3.4 Mpc, corresponds to an absolute magnitude $B=-5.6$, consistent
with a $60\msun$ MS star of spectral type O5V or a B2Ib supergiant
with mass $\sim 20-25\msun$.

\end{itemize}

\section{Intrinsic optical emission from massive donors in binaries}

In the following we present the evolutionary tracks of binary systems
with high mass companion stars and stellar-through-intermediate mass
BHs. The results of this computation will be compared with the
observed properties of ULX systems in the next Section. The
calculation has been carried out
adopting the model described in Section 2 and using a modified version
of the Eggleton code \citep{patruno}.
We focus on the modeled optical properties of massive ($10$--$50
M_\odot$) donor stars in binaries during mass transfer phases. The
accreting black hole masses are in the range 10--500
$M_\odot$. Binaries with less massive donors ($\sim 2$--$10 M_\odot$) and
stellar mass black holes have been investigated by
\cite{2005MNRAS.356..401R}, while calculations of the evolutionary
tracks for $1-9 M_\odot$ stars accreting onto stellar BHs and IMBHs are presently
under way.

As shown by \cite{madetal06}, who evolved $\sim 10^5$ BH binaries to
study the production efficiency of ULX systems, the initial orbital
period for triggering a RLOF phase during MS and originating a
significantly large population of active ULXs in binaries with massive
donors
is in the range $\sim 1$--$5$ days. We therefore limit our analysis to
a selected number of models with initial parameters having the values
reported in Table 2. The three groups (A, B and C) are characterized
by the different evolutionary phase at which mass transfer sets in
(see also \S 2.1). For Case A, the bolometric luminosity of the system
reaches at least $10^{40}$-$10^{41}$ erg s$^{-1}$ (depending on the
donor mass) at some stage during MS. For stellar mass BHs, a suitable
degree of beaming needs to be invoked in order to reach similar
isotropic luminosities.

\begin{table*}
\caption{Initial parameters for ULX binary systems. Case A refers to binaries where the donor starts the 
first Roche lobe contact phase during MS, while case B and C refer to donors starting RLOF 
during the H-shell or He-shell burning phase, respectively. The 
first and second columns are the initial donor and accretor masses, while the third column is the initial 
orbital period of the binary system.}
\label{tab2}
\begin{tabular}{llll}
\hline\hline
Case  & M($\msun$) & $M_{BH}(\msun)$ & Initial $P_{orb}$ (days) \\ 
\hline
A & 10,15,30,50 & 10,100,500  & 1-2  \\
B & 10,15,30,50 & 10,100,500  & 5-15,400-500   \\
C &  10,15,30,50 & 10,100,500 & 800-1000 \\
\hline 
\end{tabular} 
\end{table*}

\begin{figure}
\centerline{ \leavevmode
\includegraphics[width=85mm]{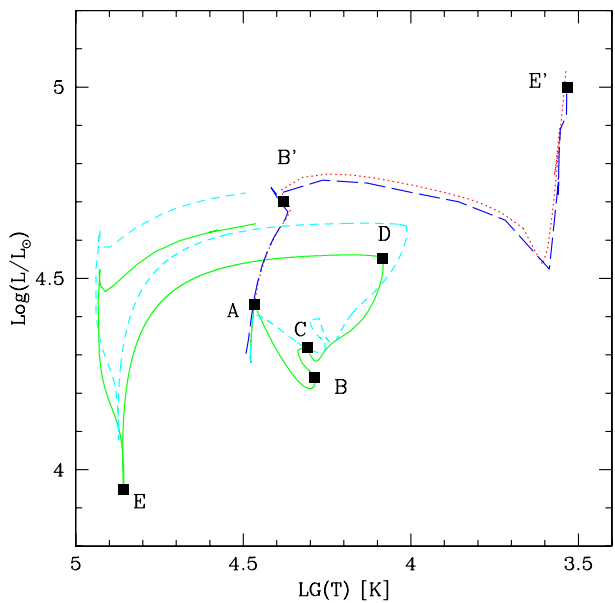}
}
\caption{HR diagram for a 15 $M_{\odot}$ donor in a binary system with
a black hole of 10 and 100 $M_{\odot}$ ({\it green-solid and light blue-dashed line})
 and evolving as a single star ({\it dark blue-long dashed line}). 
The {\it red-dotted line}
is the Geneva track and has been plotted for comparison. During the
binary evolution RLOF sets in during MS (point A) and originates a qualitatively
different track caused by the donor loosing its envelope onto the BH.
Two episodes of mass transfer occur (segments A--B and C--D). 
The point B' refers to the TAMS and is in the equivalent position 
of point B in the binary track.
In point E and E' the donor 
ignites the He-core. The tracks end with the He-core ignition for single 
stars and with the C-core ignition for binary systems.
}
\label{fig1}
\end{figure}

In Figure~\ref{fig1} we show the evolutionary tracks on the HR diagram
of single stars and stars accreting through RLOF in a binary system.
As we want to show the differences between the evolution of a single
star and that in an accreting binary system, we do not include
irradiation. 
The donor and the single star tracks refers to a 15$\msun$ star,
whereas the accretors are a $10\msun$ BH and an IMBH of $100\msun$.
When RLOF sets in (point A), the track of
a star in a binary starts to diverge from that of a single star. For
the models shown in Figure~\ref{fig1}, this occurs during MS. The
luminosity starts to decrease with time as the decrease in surface
temperature is not accompanied by a corresponding increase in radius
(which is limited by the Roche lobe). Thus, during MS (segment A--B),
a single star is generally more luminous than a star of the same mass
in a binary system. The temperature and luminosity of a MS star in a
binary system are more similar to those of a less massive star. This
may lead to misinterpret a real MS star with one that is crossing the
Hertzsprung Gap, causing uncertainties (up to $2-5\msun$ for a
15$M_{\odot}$ star) in the estimate of the donor mass.  The system
detaches in point B after an accretion phase of $\sim 10^{7}\rm\,yrs$,
very close to the terminal age main sequence.

During the H-shell burning the envelope of the star expands and the
binary starts a new episode of mass transfer lasting
$10^{4}$--$10^{5}\rm\,yrs$ (segment C--D). The system widens because
of the large amount of material that is rapidly transferred from the
secondary to the BH. As a consequence, the RL increases significantly
and, despite the star becomes cooler, the luminosity increases. While
during this phase single stars simply move to the right of the HR
diagram, becoming progressively colder and undergoing core contraction
and envelope expansion (Hertzsprung Gap and Giant branch), stars in
binary systems do not cool so much and show a lower luminosity (by a
few tenth of magnitude) than single stars of the same mass at
a corresponding evolutionary phase. 
At a certain phase during H-shell burning (point D), all the H
envelope is accreted and what is left of the star (the helium core)
contracts causing the binary to detach. From point D to carbon core
ignition, the donor remains always in a detached phase.
At point E the He-core ignites ($L_{He}/L_{H}>10$) and the evolution
becomes extremely fast with the track stopping after the onset of
C-core burning.

\begin{figure}
\centerline{ \leavevmode
\includegraphics[width=85mm]{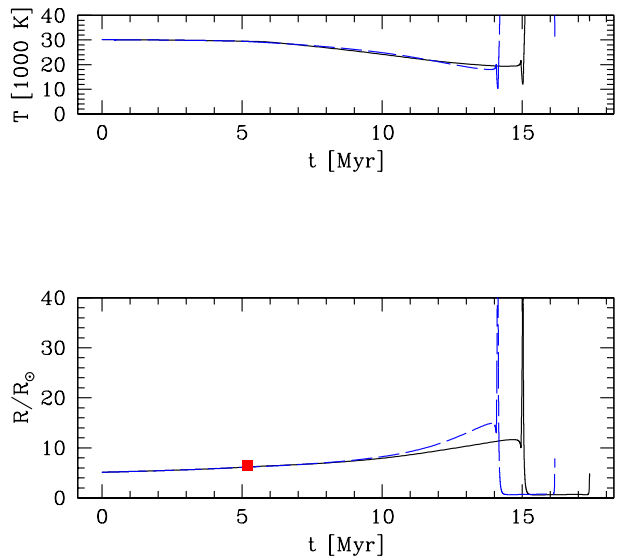}
}
\caption{Evolution of surface temperature and radius for a 15$\msun$ donor around 
a 10$\msun$ ({\it black-solid line}) and a 100$\msun$ ({\it
blue-dashed line}) black hole. The red square marks the onset of the
contact phase during MS. The two binaries have the same parameters of
those shown in Figure~\ref{fig1}. We note that the lifetime of a donor
undergoing a case A mass transfer is slightly longer than the MS
lifetime of the corresponding single star ($10^{6}-10^{7}\rm\,yrs$)
because the nuclear burning timescale is slightly increased as a
consequence of the mass lost during mass transfer. }
\label{fig2A}
\end{figure}

Figure~\ref{fig1} shows also the evolutionary track for a donor
accreting onto a $100 M_\odot$ BH starting the contact phase when the
radius of the companion is comparable to the $10 M_\odot$ BH
case.
The overall evolution is similar to the stellar mass BHs, apart
from the fact that stars accreting onto an IMBH are more luminous.
This is a consequence of the fact that, while the surface temperature
remains essentially comparable at a corresponding evolutionary phase,
the RL of the donor (and therefore its radius during the contact
phase) grows more rapidly in an IMBH binary then in a stellar BH
system (see Figure~\ref{fig2A}). 

\begin{figure}
\centerline{ \leavevmode
\includegraphics[width=85mm]{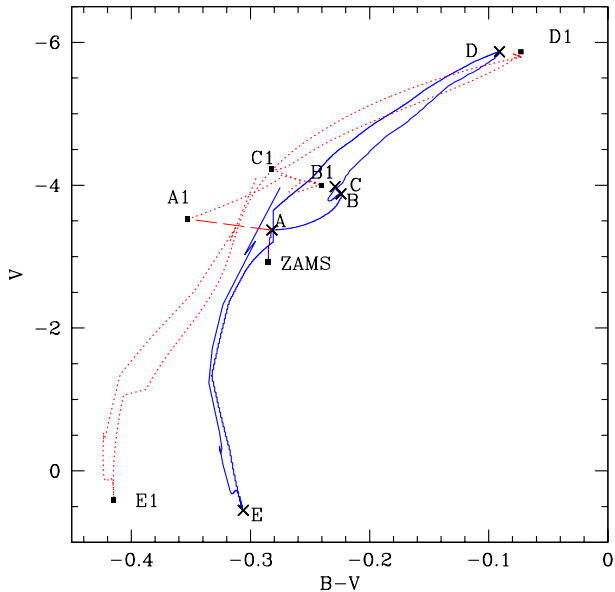}
}
\caption{CM diagram for a 15 $M_{\odot}$ donor accreting on a 10 $M_{\odot}$ 
BH. The {\it blue-solid line} is the CM diagram for a non-irradiated
star, while the {\it red-dotted line} takes into account both the
X-ray irradiation of the stellar surface and the optical contribution
of the accretion disc. The two tracks share the initial part of the
curve (from ZAMS to point A), as no irradiation is present at this
stage. Mass transfer sets in at point A.  The contact phase ends in
point B (and B1), very close to the terminal age main sequence. In
point C (C1) the star is burning its H-shell and a new contact phase
begins lasting up to point D (D1).  In point E (E1) the star begins He
core burning. After point D (D1) no other contact episodes occur
throughout the evolution.}

\label{fig2}
\end{figure}

\begin{figure}
\centerline{ \leavevmode
\includegraphics[width=85mm]{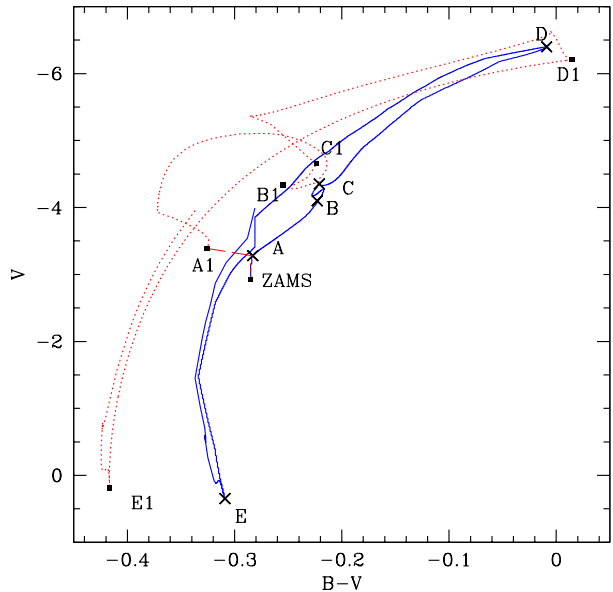}
}
\caption{CM diagram for a 15 $M_{\odot}$ donor accreting on a 100 $M_{\odot}$ 
BH. The labels have the same meaning as in fig.~\ref{fig2}. The irradiation
has a stronger effect than in the stellar BH case, giving differences in magnitudes as high as 0.2 mag on the B-V scale during the main sequence. }

\label{fig2B}
\end{figure}

An important observational diagnostic for stars is represented by the
color-magnitude (CM) diagram. The CM diagram of a 15 $M_{\odot}$ donor
accreting onto a 10 and 100 $M_{\odot}$ BH is reported in
Figure~\ref{fig2} and Figure~\ref{fig2B}. 
The tracks calculated for single stars differ in a
relevant way with respect to those of a Case A binary that starts mass
transfer during MS, as the cold branch that is covered in the CM
diagram during the post giant phase is absent. The notation is the
same used in the HR diagram.
Therefore, the contact phases correspond to the segments A--B (MS) and
C--D (H-shell burning). After point D the system detaches and the star
turns back on the CM diagram, following backwards the same path
crossed during the H-shell contact phase and the MS.  This behavior
is strictly related to the evolution of the donor after MS discussed
above.

Figure~\ref{fig2} and Figure~\ref{fig2B} also show 
how the tracks modify after including the
optical emission of the X-ray irradiated accretion disc and donor
surface. When mass transfer sets in at point A, the evolution of the
irradiated system detaches from the non-irradiated one, as irradiation
is enhancing both the observed star luminosity and effective
temperature. For a large part of the evolution, at corresponding
evolutionary times, the track of the X-ray irradiated system is
shifted towards the blue by $\sim 0.1-0.2$ magnitudes, while showing a
comparable increase in the V magnitude. Only when, close to the end of
the second mass transfer episode (point D in Figures 1 and 2), the
system is very wide and starts to detach, irradiation becomes
negligible and the donor luminosity dominates, making the tracks
essentially coincident. After point D the absence of any contact phase
prevents the occurrence of bright X-ray emission as only a very weak
WFA stage is possible.
Therefore, after point D the two tracks coincide. For illustrative
purposes alone, in order to show the maximum contribution induced by
X-ray irradiation, we assume that all the donor wind is somehow
accreted. So, the second track in Fig.~\ref{fig2} and ~\ref{fig2B}
accounts for X-ray illumination also after point D1.


\section{Properties of the ULX counterparts}

The calculations presented in the previous Section can be used to
constrain the properties of the donor stars in ULX systems by
comparing their position on the HR or CM diagrams with the
evolutionary tracks of massive BH binaries. This approach may actually
provide interesting clues also on the properties of the binary system
itself, including the BH mass.

\subsection{Case A mass transfer}

In Figures~\ref{fig3} and~\ref{fig4} we show the position in the CM
diagram of the proposed counterparts of the four ULXs considered,
along with the evolutionary tracks of irradiated donor stars in binary
systems undergoing a case A mass transfer with optical contamination
from the disc. The tracks are plotted only during the contact phases
(e.g. A-B, C-D) as only during these phases, the ULX is active. 
In the stellar mass BH case, we assume that the total isotropic luminosity is
Eddington-limited. If the accretion rate becomes super-Eddington, 
we set ${\dot M}={\dot M}_{Edd}$ and the excess mass is expelled from the system.
We plotted for comparison also the track of a non-irradiated, single star of
15$M_{\odot}$.

\subsubsection{Differences between stellar BH and IMBH systems}

Comparing Fig.~\ref{fig3} and \ref{fig4} it is readily clear that, at
corresponding stages, the V magnitude of an IMBH binary is smaller by
$\sim$0.5 mag (i.e. the luminosity higher by a factor $\sim$1.5) than
that of a stellar BH binary. The main reason for this difference stems
from the larger contribution to the optical emission coming from the
irradiated accretion disc around an IMBH and from the intrinsic
larger luminosities of donors around IMBH (see Fig.\ref{fig1}). 
Assuming a fixed orbital
separation and a standard tidal truncation scenario with the accretion
disc radius $R_{D}\sim70\%R_{L}$, two binaries with mass ratios
$q=0.1$ and $q=1$ have $R_{D,q=0.1}/R_{D,q=1}\sim 2$. Since for an
IMBH binary $q\simless 0.1$, whereas a stellar BH binary has $q\sim
1$, disc contamination will always be larger in the IMBH
case. Evolving a binary with a larger IMBH (500$\msun$), we found only
second order differences with respect to the tracks of a binary with a
100$\msun$ IMBH, as the Roche lobe radius is nearly the same.

A major difference between the stellar BH and IMBH concerns the
possible presence of beamed emission. In order to emit apparent
isotropic luminosities of $\simless 10^{40}$ erg s$^{-1}$ without
seriously violating the Eddington limit, a stellar mass BH binary must
find a way to collimate its emission. For a beaming factor $\sim$0.1,
both the disc and donor irradiation are strongly reduced as the X-rays
can not intercept the outer disc and stellar surfaces. Then, the
tracks on a CM diagram will be more similar to those of a
non-irradiated system. However, as shown in Fig.\ref{fig2}, the
differences in the colors and magnitudes between an irradiated and a
non-irradiated system during the contact phases are small ($\sim 0.1$
mag for the stellar BH and $\sim 0.2$mag for the IMBH case) and are therefore 
neglected considering the present photometric accuracy .

\subsubsection{Optical identification}

\begin{figure}
\centerline{ \leavevmode
\includegraphics[width=85mm]{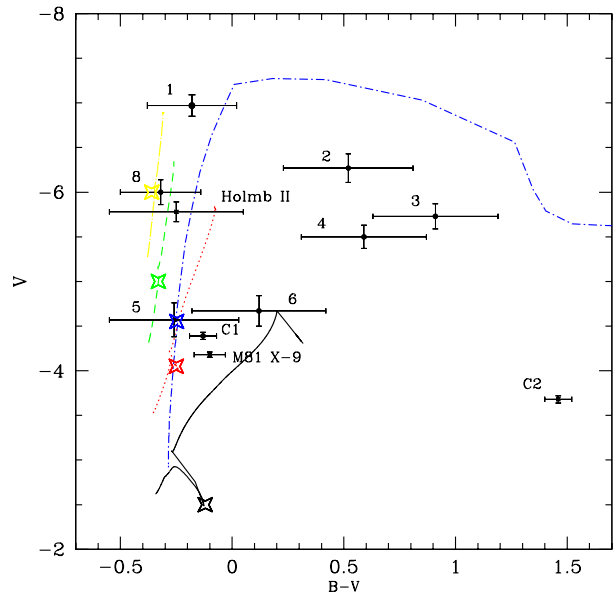}
}
\caption{
CM diagram for stars of 10, 15, 30 and 50$M_{\odot}$ ({\it
black-solid, red-dotted, green-dashed, yellow-dot dashed line},
respectively) starting mass transfer onto a 10 $M_{\odot}$ BH during
MS (case A).  The {\it blue-dot dashed line} is the track of a single
non-irradiated 15$M_{\odot}$, plotted for comparison. The crosses are
the photometric points of the counterparts of NGC 4559 X-7 (1--6, 8),
NGC 1313 X-2 (C1, C2), Holmberg II X-1 (Holm II) and M 81 X-9. The
binary tracks are plotted only during the contact phases. The starred
symbol denotes the point where the MS ends and H-shell burning sets
in.
 }
\label{fig3}
\end{figure}

\begin{figure}
\centerline{ \leavevmode
\includegraphics[width=85mm]{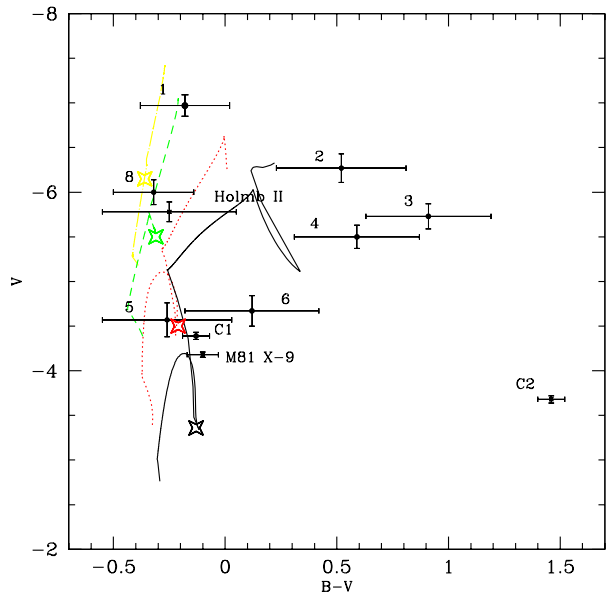}
}
\caption{
Same as Figure~\ref{fig3} for a BH of 100 $M_{\odot}$.}
\label{fig4}
\end{figure}

\textbf{NGC4559 X-7}: as can be seen from Figures \ref{fig3} and \ref{fig4}, because 
of the turnoff occurring during the giant phase, the candidate
counterparts of NGC4559 X-7 labeled 2, 3 and 4 can be immediately
ruled out while, for a stellar mass BH, object 6 is consistent with
the colors and luminosity of the least massive donor considered ($\sim
10 M_\odot$). On the other hand, the point representing objects 1, 5
and 8 overlap the evolutionary tracks of stars in binaries with mass
transfer. 
A MS star with a mass up to 50$\mdot$ is not consistent with the
colors and luminosity of object 1, regardless of the BH mass. However,
during the H-shell burning phase, a $\sim$50$\msun$ star in a stellar
mass BH system or a $\sim 30-50 \msun$ star in an IMBH system turn out
to be compatible with the photometric properties of object
1. Considering the other proposed counterparts, a MS star with a mass
smaller than 15$M_{\odot}$ around a stellar BH can not account for any
of them as, at the end of the MS, the V band magnitude is too
small. Object 5 can be well explained with a MS donor of
$15$--$30\msun$ or a giant of $10$--$15\msun$.\\
\textbf{NGC1313 X-2}: considering the second contact phase in a stellar BH binary 
(beyond the starred symbol in Fig.~\ref{fig3}), we find that a donor
of $\sim$15$\msun$ undergoing H-shell burning can account for the
colors and magnitude of the candidate counterpart C1. In case of a
IMBH binary, Fig.~\ref{fig4} shows that a 10-15$\msun$ donor is
consistent with the properties of object C1 both on the MS and during
H-shell burning. On the other hand, object C2 can be firmly ruled out
as possible counterpart of the ULX.
\\
\textbf{Holmbgerg II X-1}:  if the accretor is a stellar BH,
then either a star with 15$<$M$<$30$\msun$ during the H-shell burning
phase or a donor with $\sim$50$\msun$ on the MS can account for the
observed photometric properties of the optical counterpart. For an
IMBH system, the situation is similar, except that also a 10$\msun$
donor is compatible when in the H-shell burning phase.\\
\textbf{M81 X-9}: the optical counterpart of this ULX
is the faintest considered in this paper and is compatible only with a
$<$15$\msun$ donor during H-shell burning for a stellar BH or with a
$\simless$10$\msun$ donor on the MS or H-shell burning phase for an
IMBH.

\subsubsection{Mass transfer rate}

\begin{figure}
\centerline{ \leavevmode
\includegraphics[width=85mm]{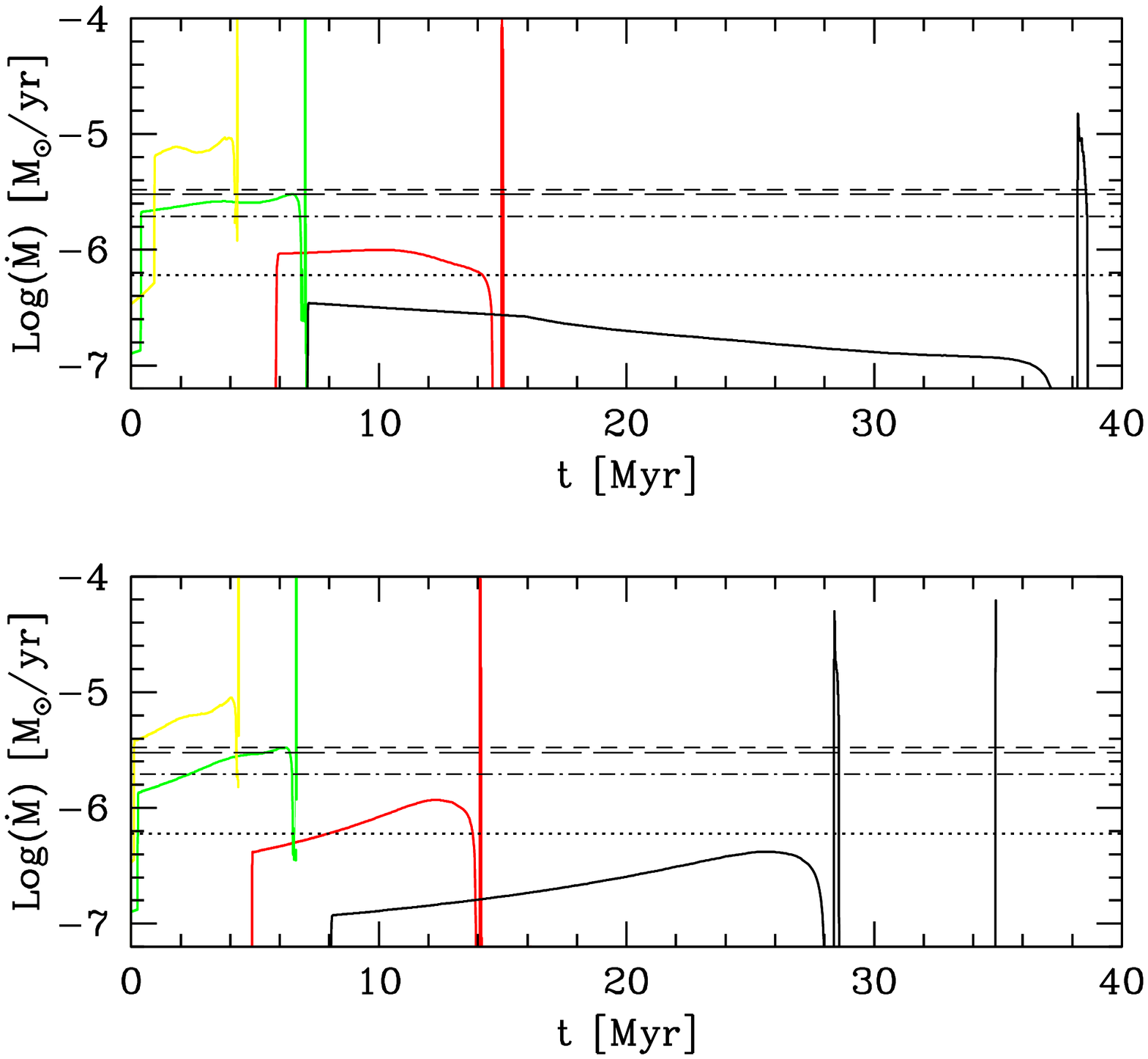}
}
\caption{Mass transfer rate of 10, 15, 30 and 50$\msun$ donors 
around a 10 $\msun$ stellar BH ({\it upper panel}) and 100$\msun$ IMBH
({\it bottom panel}) undergoing a case A mass transfer. The mean mass
transfer rate required to produce the X-ray luminosities reported in
Table \ref{tab1} is indicated with a {\it dotted} (NGC1313 X-2) {\it
dot-dashed} (M81 X-9), {\it long dashed} (Holmberg II) and {\it short
dashed} (NGC4559 X-7) line. The adopted efficiency of conversion of
gravitational potential energy into radiation is $\eta=0.1$.  }
\label{fig7}
\end{figure}

In Figure~\ref{fig7} we show the mass transfer rate in binary systems
undergoing a case A mass transfer onto a stellar BH of $10\msun$ or an
IMBH of $100\msun$. The horizontal lines are the X-ray luminosities of
the four ULXs reported in Table \ref{tab1} and, as already discussed
in \S 3, should be considered as lower limits for the bolometric
luminosity of each source (which is estimated to be a factor $\sim 2$
higher; see Table \ref{tab1}). We note that these values are
indicative as both the sources show a factor of a few variability and
the measurements of the X-ray fluxes are usually affected by rather
large errors.\\ 
In the \textbf{stellar BH case}, the mass transfer rate does not provide any
further constraint as the relation between mass loss and
luminosity becomes $L_{\rm\, bol}=\dot{M}\eta\rm\, c^{2}\rm\, b$ where
$\rm\,b$ is the beaming factor whose value is unknown .
In the present paper we are considering only standard geometrically 
thin and optically thick discs (Shakura-Sunyaev discs) 
and any genuine isotropic super-Eddington luminosity produced 
by alternative radiation-pressure dominated discs (e.g.~\citealt{b02},~\citealt{b06})
will give a different optical contamination with respect to that 
calculated in the present work. \\ 
In the \textbf{IMBH case}, the mass transfer rates are very similar to those
of the stellar BH binaries. However we can further constrain the donor
masses for NGC4559 X-7, as it is possible to rule out all the MS
donors with $M\simless$30$\msun$. For NGC1313 X-2 a $\sim 15\msun$
donor can give the required $\dot{M}$ during a certain fraction of its
MS lifetime. For Holmberg II X-1 no further constraints can be
obtained, while for M81 X-9 it is possible to rule out the
10$M_{\odot}$ MS donor, leaving as unique possibility a 10$\msun$
donor in the H-shell burning phase.\\ In table~\ref{tab3} and
~\ref{tab4} we show all the binaries compatible with the observed
optical counterparts given the combined constraints coming from the
mass transfer rate and the CM diagram.

\begin{table*}
\begin{center}
\caption{Consistent donor masses for case A mass transfer onto a $10\msun$ 
black hole. A 'X' (or the name of the donor when there are multiple
optical counterparts for the same ULX) marks the entry corresponding
to a certain donor mass and contact phase (MS or H-shell). Each donor
must satisfy the condition that the track on the CM diagram crosses the
optical counterpart.}
\begin{tabular}{c|c|c|c|c|c|c|c|c}\label{tab3}
&\multicolumn{2}{|c|}{10$\msun$}&\multicolumn{2}{c|}{15$\msun$}&\multicolumn{2}{|c|}{30$\msun$}&\multicolumn{2}{c|}{50$\msun$}\\ 
\cline{2-9}
   & MS & H-shell & MS & H-shell & MS & H-shell & MS & H-shell \\
\hline\hline
NGC4559 X-7  & & 6 & & 5 & 5 & 8 & 8 & 1  \\
\hline
NGC1313 X-2 & & & & C1 & & & &    \\
\hline
Holmberg II X-1 & & & & X & & X & X & \\
\hline
M81 X-9  & & X & & & & & & \\
\hline 
\end{tabular} 
\end{center}
\end{table*}
 \begin{table*}
\begin{center}
\caption{As Table~\ref{tab3} but for a 100$\msun$ IMBH and with the 
further constraint on the mass transfer rate. Each donor in the table 
has a mass transfer rate high enough to reach at least the observed X-ray luminosity assuming isotropic emission.}
\begin{tabular}{c|c|c|c|c|c|c|c|c}\label{tab4}
&\multicolumn{2}{|c|}{10$\msun$}&\multicolumn{2}{c|}{15$\msun$}&\multicolumn{2}{|c|}{30$\msun$}&\multicolumn{2}{c|}{50$\msun$}\\ 
\cline{2-9}
   & MS & H-shell & MS & H-shell & MS & H-shell & MS & H-shell \\
\hline\hline
NGC4559 X-7  & & 5 & & 5,8 &  & 1,8 & 8 & 1  \\
\hline
NGC1313 X-2 & & C1 & C1 & C1 & & & &    \\
\hline
Holmberg II X-1 & & X & & X & & X & X & \\
\hline
M81 X-9  & & X & & & & & & \\
\hline 
\end{tabular} 
\end{center}
\end{table*}


\subsection{Case B and C mass transfer}

When the donor star starts a case B or C mass transfer, the evolution
of the binary may be completely different with respect to what
reported above. If the contact phase sets in at the beginning of the
Hertzsprung Gap, when the donor radius is still relatively small
(e.g. $\rm\,P\simless 10\rm\,days$ for a 15$M_{\odot}$ donor), the
contact phase is extremely short lived (a few $10^{4}\rm\,yrs$) and
$\dot{M}$ can reach a few $10^{-5}\msun\rm\,yr^{-1}$ (in some cases up
to $10^{-3}\msun\rm\,yr^{-1}$). The tracks on the CM diagram are
similar to those calculated for case A, with a turnoff during the
H-shell burning at the point where the system becomes detached. If the
contact phase sets in at the end of the Hertzsprung Gap ($P\simgreat
10\rm\,days$ for a 15$M_{\odot}$ donor) or later (case C mass
transfer), $\dot{M}$ is comparable to that of a case A mass transfer
binary during H-shell burning, but now the donor enters the region of
the single giant stars on the CM diagram and the differences with the
case A tracks become sharp.

In this situation all the previously excluded optical counterparts of
NGC4559 X-7 (objects 2, 3 and 4) become good candidates, while objects
1, 5, 6 and 8 are ruled out. In the case of NGC 1313 X-2, both
counterparts can be excluded. In particular, object C2 can be
definitively ruled out as its position is not consistent with any
irradiated or non-irradiated track. Also the counterparts of Holmberg
II X-1 and M81 X-9 are not consistent with any case B or C binary
system. Finally, the most massive donors (M$\greateq$30$\msun$) are
too bright at this stage to be proposed as candidate counterparts
regardless of the compact object mass.

\section{Discussion and Conclusions}

In this paper we have shown that the effects of the binary evolution
on the optical properties of a donor star in a ULX binary system can
significantly alter its colors and evolution. We included also the
contribution of the optical emission of the accretion disc and the
X-ray irradiation of the donor and disc surfaces. The evolutionary
track of a donor in an accreting binary on the CM diagram is very
different with respect to that of a single isolated star. These
important differences can not be overlooked when trying to identify
the donor mass of a ULX. We calculated tracks for stellar and
intermediate mass black holes, and demonstrate the brighter nature of
the IMBH as an intrinsic phenomenon produced by the low mass ratio in
the binary.

The photometric data of four ULX counterparts have been compared with
the evolutionary tracks of massive donors on the CM diagram. We find
that the counterparts 2, 3 and 4 of NGC 4559 X-7 are consistent only
with donors of $10$--$15\msun$ undergoing a case B or C mass transfer
episode.
The only possibility for objects 5 and 8 is a very massive
($\sim50\msun$) MS donor or a H-shell burning donor with mass between
10-15 and 30$\msun$. Finally, object 1 is compatible only with a very
massive ($M\simgreat 50\msun$) donor in a H-shell burning phase. Our
results are quite different with respect to what obtained in previous
analysis. \cite{scpmw04} compared theoretical evolutionary tracks of
single stars between 9 and 25$\msun$ with the position in the CM
diagram of the counterparts of NGC4559 X-7. They find that the colors
and luminosity of six out of seven stars are consistent with the
tracks of main sequence, blue or red supergiant with masses in the
range 10-15 $M_\odot$ and ages $\sim 20$ Myr. Their favored
counterpart, object 1, was identified with a main sequence or blue
supergiant of $\sim 20 M_\odot$ and an age of $\sim 10$ Myr, although
\cite{scpmw04} recognize that X-ray irradiation may affect its colors
and hence its classification. Irradiation effects were taken into
account by \cite{cpw07}, giving a donor mass of 5-20$M_{\odot}$ for
object $1$. In their work, however, \cite{cpw07} considered
irradiation on single stars and did not take into account the effects
of the markedly different evolution of a donor star in a binary
system.

As far as NGC1313 X-2 is concerned, we definitely rule out the
candidate counterpart C2 and identify C1 as a H-shell burning donor of
10$\msun<$M$<$15$\msun$ around a stellar mass black hole or a
$\simless 15 \msun$ donor undergoing RLOF during MS (or the H-shell
burning phase) in an IMBH binary system. On statistical grounds alone,
it is then more likely that NGC1313 X-2 hosts a $\sim 100 \msun$ BH
rather than a stellar mass BH as the H-shell burning phase is much
shorter than the MS phase. Our result for object C2 leaves only C1 as
the likely counterpart of NGC 1313 X-2, in agreement with the evidence
coming from the refined X-ray astrometry of the field recently
reported by \cite{Liu07}.  On the basis of single star isochrone
fitting of the parent stellar population, \cite{pak06} and
\cite{ramsey06} estimate a maximum MS mass of 8--9$\msun$ for object
C1. \cite{Liu07} find consistency with either a $\sim 8\msun$ star of
very low metallicity or an O spectral type, solar metellicity star of
$\sim 30\msun$. Our estimated range of donor masses for object C1
appears more in agreement with the value (10-18$\msun$) reported by
\cite{mucetal07}, probably because they also included the contribution
of the disc and irradiation effects.
In Holmberg II X-1 there are several possibilities, and we can only
rule out stars with M$<$10$\msun$ (\citealt{cpw07} give a lower bound
of 5$\msun$), while for M81 X-9 we are left with the unique
possibility that the donor is a $\sim 10\msun$ star in the H-shell
burning phase.

We conclude therefore that in all previous work where the binary
evolution and/or irradiation effects ware not taken into account, the
donor mass has been systematically overestimated. We find that mass
transfer occurring at late stages is very unlikely, not only for the
short timescale of the contact phase ($10^{3}$--$10^{5}$yrs) but also
for the incompatibility of the optical colors with the majority of the
observed counterparts. If mass transfer sets in during MS, two contact
phases occur (segments A--B and C--D). The mass transfer timescales
are $t_{A-B}\sim 10^{6}-10^{7}\rm\,yrs$ and $t_{C-D}\sim
10^{3}-10^{5}\rm\,yrs$ (depending on the donor mass). Therefore
assuming a flat distribution of periods between 1 and $\sim$10 days, as
during MS the contact phase is reached for $P_{orb}\simless 2$days, we
expect a factor $\sim \frac{10}{2}\times \frac{t_{A-B}}{t_{C-D}}\simeq
500$--$5000$ more main sequence systems than H-shell burning donors.

\section{Acknowledgements}

We would like to thank Onno Pols and Jasinta Dewi for the kind support
in the use of the Eggleton code, and Simon Portegies Zwart, Monica
Colpi and Manfred Pakull for useful discussions. We
also thank the referee of a previously submitted version of this work
for his valuable comments that helped to improve our paper.

\newcommand{\nat}{Nat}
\newcommand{\mnras}{MNRAS}
\newcommand{\aj}{AJ}
\newcommand{\pasp}{PASP}
\newcommand{\aap}{A\&A}
\newcommand{\apjl}{ApJ}
\newcommand{\apss}{ApSS}
\newcommand{\apjs}{ApJS}
\newcommand{\aaps}{AAPS}

\end{document}